# Conservative–dissipative forces and heating mediated by fluctuation electromagnetic field: two plates in relative nonrelativistic motion

*G.V. Dedkov, A.A. Kyasov*

Nanoscale Physics Group, Kabardino –Balkarian State University, Nalchik, 360004, Russian Federation. E-mail: gv_dedkov@mail.ru

**Abstract** –We calculate heating rate, attractive conservative and tangential dissipative fluctuation electromagnetic forces felt by a thick plate moving with nonrelativistic velocity parallel to a closely spaced another plate in rest using relativistic fluctuation electrodynamics. We argue that recently developed relativistic out of equilibrium theory of fluctuation electromagnetic interactions (Volokitin et. al., Phys.Rev. B78, 155437 (2008)) has serious drawbacks.

## 1.Introduction

Vacuum attraction, friction and heat exchange of neutral nonmagnetic bodies moving with relative velocity $V$ are the well known effects of electromagnetic fluctuations. To date, however, theoretical description of many aspects of fluctuation electromagnetic interactions (FEI) has encountered with a lot of problems attracting growing attention (see the reviewing papers [1-8]). Of these one can mention the problems of dissipative (frictional) forces [2, 6], thermodynamics puzzles of the Lifshitz theory [1,4,7] and non –equilibrium Casimir forces [9,10] (see also references there in), etc. The range of applications involving FEI is very wide and extends from atomic physics and elementary particle physics to astrophysics and cosmology. By measuring Casimir forces, for example, one can study structure of quantum vacuum and determine restrictions on the magnitude of hypothetical long-range forces that are corrections to Newtonian gravitational forces [1]. Under outer space conditions, FEI between dust particles and background electromagnetic radiation can play an important role in evolution of gas –dust clouds.

In general, FEI is associated with quantum and thermal fluctuations in the polarization and magnetization of condensed bodies. Calculating the spectrum of electromagnetic fluctuations for arbitrary geometry of interacting bodies faces severe difficulties. By this reason, exact analytical or numerical solutions of the problems involving FEI (if these solutions exist) are of paramount physical importance [9]. One problem of that kind has been recently solved and reviewed in detail in our papers [6,8]. It corresponds to the geometrical configuration "small spherical particle –plate", referred to as configuration "1" (see Fig.1(a)). But historically, since pioneering works by Casimir and Lifshitz [11,12] to calculate FEI, the most widely used configuration has been regarded another one, corresponding to smooth featureless parallel plates divided by a vacuum gap of width $z$ (Fig.1(b)) [1,3-5,10,12-15]. In what follows this configuration is referred to "2"[1].

It is pertinent to note that, contrary to configuration 1, a strict relativistic solution of the problem FEI in configuration 2 out of thermal equilibrium is not yet obtained. Different aspects of this matter have been previously discussed in [2,5] and revealed many contradictions between results of several authors. Moreover, even in the simplest nonrelativistic case, the problem still turns out to be insufficiently clear [16,17]. As a matter of fact, configuration 1 is being considered by many authors as less important, secondary one, because the Casimir–Polder force between a small particle (an atom) and a plate can be calculated in the limit of rarified matter for one of the interacting plates using the Lifshitz formula for the Casimir force between two parallel plates [12-15]. However, as we have recently shown [17-19], a net correspondence between both configurations 1 and 2 in dynamic and out of equilibrium situations is not trivial. Thus, in our opinion, an independent significance of configuration 1 in the theory of FEI has not been properly appreciated.

The aim of this paper is to work out several drawbacks in configuration 2 using our exact solution in configuration 1. We obtain retarded expressions for the conservative –dissipative forces and rate of heat exchange at nonrelativistic relative velocity of the plates. Following [16,17], in order to get these results, we put forward a correspondence rule between configurations 1,2. We also critically discuss the recently proposed relativistic theories of FEI [20,21]. We argue that rigorous solution of the general relativistic problem in configuration 2 still presents a challenge for further investigation.

## 2. Basic relativistic results in configuration 1

We begin with the exact relativistic expressions which we obtained in configuration 1 for the conservative (dissipative) forces and rate of heating caused by FEI [6,8] (see Eqs.(1)-(3)). A particle is modeled by a sphere of radius $R$, and the dipole approximation $R/z \ll 1$ is assumed, with $z$ being the distance between the center of the sphere and the plate. Geometry of motion of the particle and the coordinate system used are shown in Fig.1(a). It is assumed that the particle has temperature $T_1$, whereas the plate and surrounding electromagnetic vacuum background are at equilibrium with temperature $T_2$. The global system of magnetodielectric bodies is out of thermal equilibrium, but in a stationary regime.

---

[1] Note that in Refs.[8,18,19] we used an opposite way of numeration : "1" denoted configuration "2" in this work and vice versa.

In writing (1)-(3), we have used Eqs. (12)-(14) from Ref. [6] for the force components $F_x, F_z$ and heating rate $dQ/dt$, which have been simplified further by expanding the integration domains over the wave vectors $k_x, k_y$ down to full axes $(-\infty, +\infty)$ [16]. A slight difference from [16] is that we still keep the terms being responsible for the particle interaction with vacuum background [6,8]. These terms do not depend of $z$ and come into expressions for $F_x$ and $dQ/dt$ (the second integral terms):

$$F_x = -\frac{\hbar \gamma}{2\pi^2} \int_0^\infty d\omega \int_{-\infty}^{+\infty} dk_x \int_{-\infty}^{+\infty} dk_y k_x \left[ \alpha_e''(\gamma \omega^+) \text{Im}\left( \frac{\exp(-2q_0 z)}{q_0} R_e(\omega, \mathbf{k}) \right) + (e \leftrightarrow m) \right] \cdot$$
$$\cdot \left[ \coth\left( \frac{\hbar \omega}{2k_B T_2} \right) - \coth\left( \frac{\gamma \hbar \omega^+}{2k_B T_1} \right) \right] -$$
$$- \frac{\hbar \gamma}{\pi c^4} \int_0^\infty d\omega \omega^4 \int_{-1}^1 dx x(1+\beta x)^2 \left[ \alpha_e''(\gamma \omega_1) + \alpha_m''(\gamma \omega_1) \right] \cdot \left[ \coth\left( \frac{\hbar \omega}{2k_B T_2} \right) - \coth\left( \frac{\gamma \hbar \omega_1}{2k_B T_1} \right) \right] \quad (1)$$

$$F_z = -\frac{\hbar \gamma}{2\pi^2} \int_0^\infty d\omega \int_{-\infty}^{+\infty} dk_x \int_{-\infty}^{+\infty} dk_y \left\{ \begin{array}{l} \alpha_e''(\gamma \omega^+) \text{Re}[\exp(-2q_0 z) R_e(\omega, \mathbf{k})] \coth\left( \frac{\gamma \hbar \omega^+}{2k_B T_1} \right) + \\ + \alpha_e'(\gamma \omega^+) \text{Im}[\exp(-2q_0 z) R_e(\omega, \mathbf{k})] \coth\left( \frac{\hbar \omega}{2k_B T_2} \right) + (e \leftrightarrow m) \end{array} \right\} \quad (2)$$

$$\frac{dQ}{dt} = \frac{\hbar \gamma}{2\pi^2} \int_0^\infty d\omega \int_{-\infty}^{+\infty} dk_x \int_{-\infty}^{+\infty} dk_y \omega^+ \cdot \left[ \alpha_e''(\gamma \omega^+) \text{Im}\left( \frac{\exp(-2q_0 z)}{q_0} R_e(\omega, \mathbf{k}) \right) + (e \leftrightarrow m) \right]$$
$$\cdot \left[ \coth\left( \frac{\hbar \omega}{2k_B T_2} \right) - \coth\left( \frac{\gamma \hbar \omega^+}{2k_B T_1} \right) \right] + \quad (3)$$
$$+ \frac{\hbar \gamma}{\pi c^3} \int_0^\infty d\omega \omega^4 \int_{-1}^1 dx(1+\beta x)^3 \left[ \alpha_e''(\gamma \omega_1) + \alpha_m''(\gamma \omega_1) \right] \cdot \left[ \coth\left( \frac{\hbar \omega}{2k_B T_2} \right) - \coth\left( \frac{\gamma \hbar \omega_1}{2k_B T_1} \right) \right]$$

$$\Delta_e(\omega) = \frac{q_0 \varepsilon(\omega) - q}{q_0 \varepsilon(\omega) + q}, \quad \Delta_m(\omega) = \frac{q_0 \mu(\omega) - q}{q_0 \mu(\omega) + q}, \quad q = \left(k^2 - (\omega^2/c^2)\varepsilon(\omega)\mu(\omega)\right)^{1/2}$$

$$q_0 = (k^2 - \omega^2/c^2)^{1/2}, \quad k^2 = |\mathbf{k}|^2 = k_x^2 + k_y^2, \quad \beta = V/c, \quad \gamma = (1-\beta^2)^{-1/2}, \qquad (4)$$

$$\omega^+ = \omega + k_x V, \quad \omega_1 = \omega(1 + \beta x)$$

$$R_e(\omega, \mathbf{k}) = \Delta_e(\omega)\left[2(k^2 - k_x^2 \beta^2)(1 - \omega^2/k^2 c^2) + (\omega^+)^2/c^2\right] +$$
$$+ \Delta_m(\omega)\left[2k_y^2 \beta^2 (1 - \omega^2/k^2 c^2) + (\omega^+)^2/c^2\right] \qquad (5)$$

$$R_m(\omega, \mathbf{k}) = \Delta_m(\omega)\left[2(k^2 - k_x^2 \beta^2)(1 - \omega^2/k^2 c^2) + (\omega^+)^2/c^2\right] +$$
$$+ \Delta_e(\omega)\left[2k_y^2 \beta^2 (1 - \omega^2/k^2 c^2) + (\omega^+)^2/c^2\right] \qquad (6)$$

In the above expressions we use currently accepted definitions $\varepsilon(\omega)$ and $\mu(\omega)$ for the frequency–dependent dielectric permittivity and magnetic permeability of the plate material, $\alpha_e(\omega)$ and $\alpha_m(\omega)$ are the frequency–dependent dipole electric and magnetic polarizabilities of the particle. One primed and double primed quantities represent the corresponding real and imaginary parts.

Despite the attraction force $F_z$ applied to the particle does not directly contain any vacuum contributions (cf. with expressions for $F_x$ and $dQ/dt$), the structure of Eq. (2) essentially depends on whether a condition of local thermal equilibrium $T_2 = T_3$ ($T_3$ is the background temperature) is fulfilled, or not [18,19] (see Fig.1(a)). Particularly, an important advantage of the relativistic problem statement in configuration 1 as compared to configuration 2 is that the presence of vacuum background has to be the basic standpoint and, correspondingly to that, we have only one large body (a thick plate) which can be in rest with respect to the background. A small particle, moving near a surface of resting plate, moves with respect to the background, as well. Correspondingly, the structure of the electromagnetic field near a plate depends on whether the system plate –background is at thermal equilibrium, or not. For configuration 2, in contrast, the problem statement in dynamic situation needs to be more elaborated even at equilibrium, $T_1 = T_2 = T_3 = T$, because only one of the plates can be in rest respectively to background, whereas another plate will be braking due to its interaction with the background (see Fig.(2(b)).Thus, in the case out of equilibrium, specifics of vacuum background makes the problem to be more complex and needs to be clarified further.

Comparing Eqs.(1)-(3) with Eqs.(12)-(14) in ref. [8], one should take into account the following analytical transformations

$$\int_{-\infty}^{+\infty}\int_{-\infty}^{+\infty} d^2k = \int_{k>\omega/c} d^2k + \int_{k<\omega/c} d^2k, \quad q_0 \to -i\tilde{q}_0, \quad q_0 = (k^2 - \omega^2/c^2)^{1/2}, \quad \tilde{q}_0 = (\omega^2/c^2 - k^2)^{1/2} \qquad (7)$$

Contrary to Eqs.(12)-(14) in Ref.[8], an important advantage of Eqs.(1)-(3) is that contributions from evanescent modes ($k > \omega/c$) and from propagating modes ($k < \omega/c$) are explicitly combined into a single integral term, since the electromagnetic modes of both types come into all resulting formulae in a quite similar way. These analytical properties are of principal importance.

**3. A system of two parallel plates in relative nonrelativistic motion: nonretarded interaction**

Up to now, transition "$2 \to 1$" from configuration 2 to configuration 1 has been employed in a routine manner by many authors to calculate the Casimir–Polder force between a resting atom and a wall [12-15]. Volokitin et. al. [5,20] have applied the same procedure to calculate friction force and heat exchange in this configuration. Transition "$2 \to 1$" is generally realized using the limit $\varepsilon_1(\omega) - 1 = 4\pi n_1 \alpha_1(\omega) \to 0$ for the material of one of the plates (the first one, for consistency), where $n_1$ and $\varepsilon_1(\omega)$ are the atomic density and dielectric permittivity of the plate material, and $\alpha_1(\omega)$ is the proper atomic dipole electric polarizability. For simplicity, in this section we assume the bodies to be made of nonmagnetic materials. As far as the conservative Casimir–Polder force is concerned, the correspondence rule reads (here and after the superscripts 1,2 discriminate configurations 1 and 2) [10]

$$F_z^{(1)}(z) = -\frac{1}{n_1 S} \frac{dF_z^{(2)}(l)}{dl}\bigg|_{l=z} \tag{8}$$

where $F_z^{(2)}(l)/S$ describes the Casimir–Lifshitz force per unit area of two parallel plates with surface area $S$ divided by a gap of width $l$. The left hand side of Eq.(8) describes the Casimir–Polder force applied to a small particle (an atom), which is located a distance $z$ apart from the plate.

Analogously to (8), the relations between the lateral forces $F_x^{(1,2)}$ and heating rates $dQ^{(1,2)}/dt$ in both configurations 1 and 2 are given by

$$F_x^{(1)}(z) = -\frac{1}{n_1 S} \frac{dF_x^{(2)}(l)}{dl}\bigg|_{l=z}, \quad dQ^{(1)}(z)/dt = -\frac{1}{n_1 S} \frac{d\dot{Q}^{(2)}(l)}{dl}\bigg|_{l=z} \tag{9}$$

It seems quite natural to believe that results obtained on using the transition "$2 \to 1$" and vice versa should be interlinked. Thus, if Eqs.(8),(9) are properly used to perform an opposite transition "$1 \to 2$", this allows to continue the solution (1)-(3) down to dynamic and non equilibrium configurations 2. This correspondence rule is taken as our guiding idea in this paper.

First, let us employ the relations (8), (9) and the correspondence rule to the simpler nonrelativistic ($\beta = V/c \to 0$) and nonretarded ($\omega z/c \to 0$) case. Then, making use of the above simplifications in (1)-(3) yields

$$F_x^{(1)}(z) = -\frac{\hbar}{\pi^2}\int_0^\infty d\omega \int_{-\infty}^{+\infty} dk_x \int_{-\infty}^{+\infty} dk_y\, kk_x \exp(-2kz)\Delta''(\omega)\alpha_e''(\omega^+)\left[\coth\left(\frac{\hbar\omega}{2k_BT_2}\right) - \coth\left(\frac{\hbar\omega^+}{2k_BT_1}\right)\right] \quad (10)$$

$$F_z^{(1)}(z) = -\frac{\hbar}{\pi^2}\int_0^\infty d\omega \int_{-\infty}^{+\infty} dk_x \int_{-\infty}^{+\infty} dk_y\, k^2 \exp(-2kz)\left[\begin{array}{l}\Delta''(\omega)\alpha_e'(\omega^+)\coth\left(\frac{\hbar\omega}{2k_BT_2}\right) + \\ + \Delta'(\omega)\alpha_e''(\omega^+)\coth\left(\frac{\hbar\omega^+}{2k_BT_1}\right)\end{array}\right] \quad (11)$$

$$\frac{dQ^{(1)}(z)}{dt} = \frac{\hbar}{\pi^2}\int_0^\infty d\omega \int_{-\infty}^{+\infty} dk_x \int_{-\infty}^{+\infty} dk_y\, k \exp(-2kz)\Delta''(\omega)\alpha_e''(\omega^+)\omega^+\left[\coth\left(\frac{\hbar\omega}{2k_BT_2}\right) - \coth\left(\frac{\hbar\omega^+}{2k_BT_1}\right)\right] \quad (12)$$

$$\Delta(\omega) = \Delta_e(\omega) = \frac{\varepsilon(\omega)-1}{\varepsilon(\omega)+1},\ \Delta_m(\omega) = 0,\ k = (k_x^2 + k_y^2)^{1/2} \quad (13)$$

Formulae (10)-(12) have been firstly derived in our papers [2,22] when solving the same nonrelativistic problem, and later in [23,24] in relativistic statement using the limit $c \to \infty$. One can see that Eqs. (10)-(12) describe contribution of evanescent modes ($k > \omega/c$) at different temperatures of the particle ($T_1$) and the sample surface ($T_2$). A contribution from propagating modes ($k < \omega/c$) in the limit $c \to \infty$ goes to zero.

Also, it is worth noticing that Eqs.(10)-(12) are valid irrespectively of the state of thermal equilibrium in the system plate –background : $T_2 = T_3$ or $T_2 \neq T_3$, where $T_3$ is the background temperature or that of distant environment bodies. The same outcome holds for the corresponding evanescent contributions in relativistic formulae (1)-(3) [6,8].

First, using Eq.(11), we write down the expression for the attraction force in configuration 1 at $V = 0, T_1 = T_2 = T$:

$$F_z^{(1)}(z) = -\frac{\hbar}{\pi^2}\int_0^\infty d\omega \int_{-\infty}^{+\infty} dk_x \int_{-\infty}^{+\infty} dk_y\, k^2 \exp(-2kz)\left[\begin{array}{l}\Delta''(\omega)\alpha_e'(\omega)\coth\left(\frac{\hbar\omega}{2k_BT}\right) + \\ + \Delta'(\omega)\alpha_e''(\omega)\coth\left(\frac{\hbar\omega}{2k_BT}\right)\end{array}\right] \quad (14)$$

Comparing (14) with (11) one sees that a transition to dynamic and thermal situations out of equilibrium is performed by transformations

$$\Delta''(\omega)\coth\left(\frac{\hbar\omega}{2k_B T}\right) \to \Delta''(\omega)\coth\left(\frac{\hbar\omega}{2k_B T_2}\right),$$

$$\alpha_e''(\omega)\coth\left(\frac{\hbar\omega}{2k_B T}\right) \to \alpha_e''(\omega^+)\coth\left(\frac{\hbar\omega^+}{2k_B T_1}\right), \quad (15)$$

$$\alpha_e'(\omega),\alpha_e''(\omega) \to \alpha_e'(\omega^+),\alpha_e''(\omega^+)$$

On the other hand, comparing (10) and (11) shows that tangential force $F_x$ is obtained from (14) on using

$$d^2 kk \to d^2 kk_x, \Delta''(\omega) \to \Delta''(\omega), \Delta'(\omega) \to \Delta''(\omega), \alpha_e'(\omega^+) \to \alpha_e''(\omega^+), \alpha_e''(\omega^+) \to -\alpha_e''(\omega^+) \quad (16)$$

Furthermore, from (10) and (12) it follows that $dQ/dt$ is obtained from $F_x$ on using

$$d^2 kk_x \to -d^2 k\omega^+ \quad (17)$$

According to the correspondence rule, since Eqs.(10)-(12) must follow from their equivalents in configuration 2 with the help of linear transformation $\varepsilon_1(\omega)-1 = 4\pi n_1 \alpha_1(\omega) \to 0$, then the quantities $F_x^{(2)}(l), F_z^{(2)}(l), \dot{Q}^{(2)}(l)$ should be related with each other similar to (15)-(17), replacing $\alpha_e(\omega) \to \Delta_1(\omega)$.

Second, let us write down the expression for nonretarded Van–der–Waals force between two parallel plates at $V=0, T_1 = T_2 = T$, which we rewrite in a more convenient form

$$F_z^{(2)}(l) = -\frac{\hbar S}{4\pi^3}\int_0^\infty d\omega \int_{-\infty}^{+\infty} dk_x \int_{-\infty}^{+\infty} dk_y k \frac{\exp(-2kl)}{\left|1-\exp(-2kl)\Delta_1(\omega)\Delta_2(\omega)\right|^2} \cdot$$
$$\cdot \left[\Delta_1''(\omega)\Delta_2'(\omega)\coth(\hbar\omega/2k_B T) + \Delta_1'(\omega)\Delta_2''(\omega)\coth(\hbar\omega/2k_B T)\right] \quad (18)$$

where $\Delta_1(\omega) = \frac{\varepsilon_1(\omega)-1}{\varepsilon_1(\omega)+1}$ and $\Delta_2(\omega) = \frac{\varepsilon_2(\omega)-1}{\varepsilon_2(\omega)+1}$, $\varepsilon_{1,2}(\omega)$ are the dielectric permittivities of the plates 1,2 respectively. Making use in (18) the transformations

$$\Delta_2''(\omega)\coth\left(\frac{\hbar\omega}{2k_B T}\right) \to \Delta_2''(\omega)\coth\left(\frac{\hbar\omega}{2k_B T_2}\right),$$

$$\Delta_1''(\omega)\coth\left(\frac{\hbar\omega}{2k_B T}\right) \to \Delta_1''(\omega^+)\coth\left(\frac{\hbar\omega^+}{2k_B T_1}\right), \quad (19)$$

$$\Delta_1'(\omega),\Delta_1''(\omega) \to \Delta_1'(\omega^+),\Delta_1''(\omega^+),$$

we immediately obtain the attraction force between two moving parallel plates out of equilibrium in configuration 2:

$$F_z^{(2)}(l) = -\frac{\hbar S}{4\pi^3}\int_0^\infty d\omega \int_{-\infty}^{+\infty} dk_x \int_{-\infty}^{+\infty} dk_y k \frac{\exp(-2kl)}{\left|1-\exp(-2kl)\Delta_1(\omega^+)\Delta_2(\omega)\right|^2} \cdot$$
$$\cdot \left[\Delta_1''(\omega^+)\Delta_2'(\omega)\coth(\hbar\omega^+/2k_BT_1) + \Delta_1'(\omega^+)\Delta_2''(\omega)\coth(\hbar\omega/2k_BT_2)\right] \quad (20)$$

Similar to (10),(11), in order to calculate $F_x^{(2)}$, one should perform in (20) the following transformations :

$$d^2kk \to d^2kk_x,\ \Delta_2''(\omega) \to \Delta_2''(\omega),\ \Delta_2'(\omega) \to \Delta_2''(\omega),\ \Delta_1'(\omega^+) \to \Delta_1''(\omega^+),\ \Delta_1''(\omega^+) \to -\Delta_1''(\omega^+) \quad (20)$$

After doing that we obtain

$$F_x^{(2)}(l) = -\frac{\hbar S}{4\pi^3}\int_0^\infty d\omega \int_{-\infty}^{+\infty} dk_x \int_{-\infty}^{+\infty} dk_y k_x \frac{\exp(-2kl)}{\left|1-\exp(-2kl)\Delta_1(\omega^+)\Delta_2(\omega)\right|^2} \Delta_1''(\omega^+)\Delta_2''(\omega) \cdot$$
$$\cdot \left[\coth(\hbar\omega/2k_BT_2) - \coth(\hbar\omega^+/2k_BT_1)\right] \quad (21)$$

Finally, making use of the transformation $d^2kk_x \to -d^2k\omega^+$ in (21), we get $\dot{Q}^{(2)}$:

$$\dot{Q}^{(2)}(l) = \frac{\hbar S}{4\pi^3}\int_0^\infty d\omega \int_{-\infty}^{+\infty} dk_x \int_{-\infty}^{+\infty} dk_y \omega^+ \frac{\exp(-2kl)}{\left|1-\exp(-2kl)\Delta_1(\omega^+)\Delta_2(\omega)\right|^2} \Delta_1''(\omega^+)\Delta_2''(\omega) \cdot$$
$$\cdot \left[\coth(\hbar\omega/2k_BT_2) - \coth(\hbar\omega^+/2k_BT_1)\right] \quad (22)$$

An important point is that the heating rates $\dot{Q}^{(1,2)}$ (Eqs.(12) and (22)) are generally appear not only due to the temperature difference between contacting bodies, but also due to a transformation of work of the lateral forces $F_x^{(1,2)}$ into heat. It is easy to show that, in the limiting case of rarified body, using Eqs.(19)-(22) with account of Eqs.(8),(9) explicitly results in Eqs.(10)-(12). The impact of magnetic properties results in additional terms (in Eqs.(20)-(22)) having the same structure and replacing $\Delta_{1,2}(\omega) = \dfrac{\varepsilon_{1,2}(\omega)-1}{\varepsilon_{1,2}(\omega)+1}$ by $\Delta_{1,2}(\omega) = \dfrac{\mu_{1,2}(\omega)-1}{\mu_{1,2}(\omega)+1}$ .

**4. A system of two parallel plates in relative nonrelativistic motion: retarded interaction**

In this section we derive more general retarded expressions for the quantities $F_x^{(2)}(l), F_z^{(2)}(l), Q^{(2)}(l)$ corresponding to two plates having arbitrary magnetodielectric properties. However, we restrict our consideration to the case of total thermal equilibrium at a temperature $T$ (see Fig.1(b). Then, substituting $T_1 = T_2 = T_3 = T$ and $\beta = V/c \to 0,\ \gamma = (1-\beta^2)^{-1/2} \to 1$ into (1)-(3) yields

$$F_x^{(1)}(z) = -\frac{\hbar}{2\pi^2}\int\limits_0^\infty d\omega \int\limits_{-\infty}^{+\infty} dk_x \int\limits_{-\infty}^{+\infty} dk_y k_x \cdot \left\{ \begin{array}{l} \alpha_e''(\omega^+)\mathrm{Im}\left[\dfrac{\exp(-2q_0 z)}{q_0}R_e(\omega,\mathbf{k})\right] \cdot \\ \cdot\left[\coth\left(\dfrac{\hbar\omega}{2k_B T}\right) - \coth\left(\dfrac{\hbar\omega^+}{2k_B T}\right)\right] + (e\to m) \end{array} \right\} \qquad (23)$$

$$F_z^{(1)}(z) = -\frac{\hbar}{2\pi^2}\int\limits_0^\infty d\omega \int\limits_{-\infty}^{+\infty} dk_x \int\limits_{-\infty}^{+\infty} dk_y \cdot$$

$$\cdot \left\{ \begin{array}{l} \alpha_e''(\omega^+)\mathrm{Re}[\exp(-2q_0 z)R_e(\omega,\mathbf{k})]\coth\left(\dfrac{\hbar\omega^+}{2k_B T}\right) + \\ + \alpha_e'(\omega^+)\mathrm{Im}[\exp(-2q_0 z)R_e(\omega,\mathbf{k})]\coth\left(\dfrac{\hbar\omega}{2k_B T}\right) + (e\leftrightarrow m) \end{array} \right\} \qquad (24)$$

$$\frac{dQ^{(1)}(z)}{dt} = \frac{\hbar}{2\pi^2}\int\limits_0^\infty d\omega \int\limits_{-\infty}^{+\infty} dk_x \int\limits_{-\infty}^{+\infty} dk_y \omega^+ \left\{ \begin{array}{l} \alpha_e''(\omega)\mathrm{Im}\left[\dfrac{\exp(-2q_0 z)}{q_0}R_e(\omega,\mathbf{k})\right] \cdot \\ \cdot\left[\coth\left(\dfrac{\hbar\omega}{2k_B T}\right) - \coth\left(\dfrac{\hbar\omega^+}{2k_B T}\right)\right] + (e\to m) \end{array} \right\} \qquad (25)$$

$$R_e(\omega,\mathbf{k}) = \Delta_{2e}(\omega)(2k^2 - \omega^2/c^2) + \Delta_{2m}(\omega)\omega^2/c^2 \qquad (26)$$

$$R_m(\omega,\mathbf{k}) = \Delta_{2m}(\omega)(2k^2 - \omega^2/c^2) + \Delta_{2e}(\omega)\omega^2/c^2 \qquad (27)$$

where the terms $(e\leftrightarrow m)$ are defined by the same expressions replacing subscripts "$e$" by "$m$". It is worth noticing that the presence of magnetic polarization terms in (23)-(25) proves to be crucially important in the following. Besides, an additional subscript "2" in (26),(27) implies that material properties must correspond to the second (resting) plate.

In order to account for the terms related with magnetic properties of the bodies, the limit of rarified body $\varepsilon_1(\omega) - 1 = 4\pi n_1 \alpha_e(\omega) \to 0$ should be supplemented by $\mu_1(\omega) - 1 = 4\pi n_1 \alpha_m(\omega) \to 0$. However, from analysis of the integrand structure in Eqs. (23)-(25) it follows that the nonretarded transition rules $\Delta_e(\omega) \to 2\pi n_1 \alpha_e(\omega)$, $\Delta_m(\omega) \to 2\pi n_1 \alpha_m(\omega)$ being used in [16], have to be modified in the form [17]:

$$\Delta_{1e}(\omega) \to \frac{\pi n_1}{q_0^2}\left[\alpha_e(\omega)\left(2k^2 - \omega^2/c^2\right) + \alpha_m(\omega)\omega^2/c^2\right] \tag{28}$$

$$\Delta_{1m}(\omega) \to \frac{\pi n_1}{q_0^2}\left[\alpha_m(\omega)\left(2k^2 - \omega^2/c^2\right) + \alpha_e(\omega)\omega^2/c^2\right] \tag{29}$$

Then, with account of (8),(9) and (23)-(29) the quantities $F_x^{(2)}(l), F_z^{(2)}(l), Q^{(2)}(l)$ take the form

$$F_x^{(2)}(l) = -\frac{\hbar S}{4\pi^3}\int_0^\infty d\omega \int_{-\infty}^{+\infty} dk_x \int_{-\infty}^{+\infty} dk_y\, k_x \cdot$$
$$\cdot\left[\frac{\mathrm{Im}\,\Delta_{1e}(\omega^+)\mathrm{Im}(\exp(-2q_0 l)\Delta_{2e}(\omega))}{\left|1 - \exp(-2q_0 l)\Delta_{1e}(\omega^+)\Delta_{2e}(\omega)\right|^2} + \frac{\mathrm{Im}\,\Delta_{1m}(\omega^+)\mathrm{Im}(\exp(-2q_0 l)\Delta_{2m}(\omega))}{\left|1 - \exp(-2q_0 l)\Delta_{1m}(\omega^+)\Delta_{2m}(\omega)\right|^2}\right] \cdot$$
$$\cdot\left[\coth\left(\frac{\hbar\omega}{2k_B T}\right) - \coth\left(\frac{\hbar\omega^+}{2k_B T}\right)\right] \tag{30}$$

$$F_z^{(2)}(l) = -\frac{\hbar S}{4\pi^3}\int_0^\infty d\omega \int_{-\infty}^{+\infty} dk_x \int_{-\infty}^{+\infty} dk_y \cdot$$
$$\cdot\left[\begin{pmatrix}\dfrac{\mathrm{Im}\,\Delta_{1e}(\omega^+)\mathrm{Re}(q_0 \exp(-2q_0 l)\Delta_{2e}(\omega))}{\left|1 - \exp(-2q_0 l)\Delta_{1e}(\omega^+)\Delta_{2e}(\omega)\right|^2}\coth\left(\dfrac{\hbar\omega^+}{2k_B T}\right) + \\ + \dfrac{\mathrm{Im}\,\Delta_{1m}(\omega^+)\mathrm{Re}(q_0 \exp(-2q_0 l)\Delta_{2m}(\omega))}{\left|1 - \exp(-2q_0 l)\Delta_{1m}(\omega^+)\Delta_{2m}(\omega)\right|^2}\coth\left(\dfrac{\hbar\omega^+}{2k_B T}\right)\end{pmatrix} + \\ + \begin{pmatrix}\dfrac{\mathrm{Re}\,\Delta_{1e}(\omega^+)\mathrm{Im}(q_0 \exp(-2q_0 l)\Delta_{2e}(\omega))}{\left|1 - \exp(-2q_0 l)\Delta_{1e}(\omega^+)\Delta_{2e}(\omega)\right|^2}\coth\left(\dfrac{\hbar\omega}{2k_B T}\right) + \\ + \dfrac{\mathrm{Re}\,\Delta_{1m}(\omega^+)\mathrm{Im}(q_0 \exp(-2q_0 l)\Delta_{2m}(\omega))}{\left|1 - \exp(-2q_0 l)\Delta_{1m}(\omega^+)\Delta_{2m}(\omega)\right|^2}\coth\left(\dfrac{\hbar\omega}{2k_B T}\right)\end{pmatrix}\right] \tag{31}$$

$$\dot{Q}^{(2)}(l) = \frac{\hbar S}{4\pi^3}\int_0^\infty d\omega \int_{-\infty}^{+\infty} dk_x \int_{-\infty}^{+\infty} dk_y\, \omega^+ \cdot$$
$$\cdot\left[\frac{\mathrm{Im}\,\Delta_{1e}(\omega^+)\mathrm{Im}(\exp(-2q_0 l)\Delta_{2e}(\omega))}{\left|1 - \exp(-2q_0 l)\Delta_{1e}(\omega^+)\Delta_{2e}(\omega)\right|^2} + \frac{\mathrm{Im}\,\Delta_{1m}(\omega^+)\mathrm{Im}(\exp(-2q_0 l)\Delta_{2m}(\omega))}{\left|1 - \exp(-2q_0 l)\Delta_{1m}(\omega^+)\Delta_{2m}(\omega)\right|^2}\right] \cdot$$
$$\cdot\left[\coth\left(\frac{\hbar\omega}{2k_B T}\right) - \coth\left(\frac{\hbar\omega^+}{2k_B T}\right)\right] \tag{32}$$

$$\Delta_{ie}(\omega) = \frac{q_0 \varepsilon_i(\omega) - q_i}{q_0 \varepsilon(\omega) + q_i},\quad \Delta_{im}(\omega) = \frac{q_0 \mu_i(\omega) - q_i}{q_0 \mu_i(\omega) + q_i},\, i = 1,2 \tag{33}$$

$$q_0 = (k^2 - \omega^2/c^2)^{1/2}, \ k^2 = |\mathbf{k}|^2 = k_x^2 + k_y^2, \ q_i = \left(k^2 - (\omega^2/c^2)\varepsilon_i(\omega)\mu_i(\omega)\right)^{1/2} \tag{34}$$

where the subscripts $i = 1,2$ numerate plates, $\varepsilon_i(\omega)$ and $\mu_i(\omega)$ are the dielectric permittivity and magnetic permeability of the involved materials. In the notations used by other authors, coefficients $\Delta_{ie}(\omega)$ and $\Delta_{im}(\omega)$ correspond to reflection amplitudes of the electromagnetic waves with $P-$ and $S-$ polarization [1,3-5,10,13-15]. At $V = 0$ Eq.(31) exactly coincides with the real frequency representation of the Casimir force in configuration 2 at equilibrium [10], while at $V \neq 0$ in the nonretarded limit (for nonmagnetic bodies) Eqs.(30)-(32) are reduced to (20)-(22).

## 5. Discussion and comparison with results of other authors

It is interesting to compare formulae (20)-(22) and (30)-(32) with results obtained by other authors. One of the first successive attempts to calculate the nonretarded dissipative force $F_x^{(2)}$ between two perfectly smooth featureless plates has been done by Pendry [25]. However, he has considered only the simplest case $T_1 = T_2 = 0$. Later, in [26] Pendry has obtained the nonretarded heating rate $\dot{Q}^{(2)}$ at $V = 0$, which expression proved to be in accordance with more general result [27] in configuration 2. Formulae (21), (22) agree with these calculations. For a review of more recent calculations see [2]. As a matter of fact, none of the known early theories did not represent all the quantities $F_x^{(2)}(l), F_z^{(2)}(l), \dot{Q}^{(2)}(l)$ as a unified set of equations similar to Eqs.(20)-(22) or (30)-(32).-

The first attempts to develop a relativistic approach for calculating the dissipative force $F_x^{(2)}$ [28- 31] turned out to be insufficiently correct, because from the results of these studies it follows that $F_x^{(2)} \to 0$ at $c \to \infty$. Further great attention to this problem and calculation of heat exchange in configuration 2 has been given by Volokitin et. al. [5,20,32]. Particularly, in their recent papers [20] the authors have presented relativistic out of thermal equilibrium theory of FEI in configuration 2. Next, based on these results and using the limit of rarified medium, they derived the expressions for $F_x^{(1)}(z), F_z^{(1)}(z), \dot{Q}^{(1)}(z)$ in configuration 1. Our results for $F_x^{(1)}(z), F_z^{(1)}(z), \dot{Q}^{(1)}(z)$ strongly differ from [20], while those for $F_x^{(2)}(l), F_z^{(2)}(l), \dot{Q}^{(2)}(l)$ are different in some crucially important points, too.

First, let us compare Eq.(30) with its counterpart, Eq.(22) in Ref. [20]. It is worth noticing that thermal state of vacuum background is not clearly defined by the authors. So, in the retarded limit at $\beta \to 0$, $T_1 = T_2 = T$, in our notations, Eq.(22) from [20] takes the form

$$F_x^{(2)}(l) = -\frac{\hbar S}{16\pi^3} \int_0^\infty d\omega \int_{k<\omega/c} d^2k \cdot k_x$$
$$\cdot \left[ \frac{\left(1-|\Delta_{1e}(\omega^+)|^2\right)\left(1-|\Delta_{2e}(\omega)|^2\right)}{\left|1-\exp(-2q_0 l)\Delta_{1e}(\omega^+)\Delta_{2e}(\omega)\right|^2} + (e \leftrightarrow m) \right] \left[ \coth\left(\frac{\hbar\omega}{2k_B T}\right) - \coth\left(\frac{\hbar\omega^+}{2k_B T}\right) \right] -$$
$$-\frac{\hbar S}{4\pi^3} \int_0^\infty d\omega \int_{k>\omega/c} d^2k \cdot k_x \exp(-2q_0 l) \cdot$$
$$\cdot \left[ \frac{\mathrm{Im}\Delta_{1e}(\omega^+)\,\mathrm{Im}(\Delta_{2e}(\omega))}{\left|1-\exp(-2q_0 l)\Delta_{1e}(\omega^+)\Delta_{2e}(\omega)\right|^2} + (e \leftrightarrow m) \right] \left[ \coth\left(\frac{\hbar\omega}{2k_B T}\right) - \coth\left(\frac{\hbar\omega^+}{2k_B T}\right) \right] \tag{35}$$

Unlike [20], in writing Eq.(35) we have assumed, for uniformity, that a moving plate is the first one (as shown in Fig.1(b)). Analysis of Eqs. (30) and (35) shows that the terms corresponding to surface evanescent modes ($k > \omega/c$) coincide with each other, whereas the terms corresponding to surface propagating modes are essentially different. Particularly, Eq.(30) does not contain the absorption coefficients $\left(1-|\Delta_{1e}(\omega^+)|^2\right)$, $\left(1-|\Delta_{2e}(\omega)|^2\right)$ etc., because the system is assumed to be in total thermal equilibrium, $T_1 = T_2 = T_3 = T$, whereas the equilibrium fluctuation electromagnetic field has the structure of oscillating , standing wave [33,34]. Moreover, the state of electromagnetic field seen in the reference frame of the moving plate should not principally differ from that one in the case of resting plate. The motion only result in the frequency shift, $\omega \to \omega^+$ and $\Delta_{1e,1m}(\omega) \to \Delta_{1e,1m}(\omega^+)$. Therefore, a physically correct solution of the problem in configuration 2 at $T_1 = T_2 = T_3 = T$ must not result in the radiation wind terms, related with the absorbtion. And what is more, if the wind terms appear in the tangential force $F_x^{(2)}$, so they do in the attraction force $F_z^{(2)}$, as well. In the last case, at $V = 0$, that has been clearly demonstrated in [9,10]. However, in Ref. [20] the wind contributions are absent, while they come into $F_x^{(2)}$ even under equilibrium conditions (see Eq.(35) and Ref. [32]).

Second, the theory [20] leads to a different temperature dependence of the Casimir forces $F_z^{(1,2)}$ at $V = 0$, as compared to Refs.[9,10] and Eqs.(2), (20). In the last case, it follows that the temperature factors of the interacting plates must come into the integrand functions in combination with the corresponding material factors, that is not the case in Ref. [20] (see Eqs.(28), (30), (31)). Besides, apart from different temperature factors, a very instructive comparison can be done between our Eq.(24) and Eq.(31) in Ref.[20], which in our notations at $\beta=0$ takes the form

$$F_z^{(1)}(z) = -\frac{\hbar}{2\pi^2} \text{Im} \int_0^\infty d\omega \int_{-\infty}^{+\infty} dk_x \int_{-\infty}^{+\infty} dk_y \cdot \exp(-2q_0 z)(k^2 - \omega^2/c^2) \cdot$$
$$\cdot [\alpha_e(\omega^+)\Delta_e(\omega) + \alpha_m(\omega^+)\Delta_m(\omega)] \cdot \left[ \coth\left(\frac{\hbar\omega^+}{2k_B T_1}\right) + \coth\left(\frac{\hbar\omega}{2k_B T_2}\right) \right]$$
(36)

Consider, for instance, the simplest case $\alpha_m(\omega)=0, V=0, T_1=T_2=T$. Then Eqs.(36) and (24) are reduced to

$$F_z^{(1)}(z) = -\frac{\hbar}{\pi^2} \text{Im} \int_0^\infty d\omega \int_{-\infty}^{+\infty} dk_x \int_{-\infty}^{+\infty} dk_y \exp(-2q_0 z)(k^2 - \omega^2/c^2)\alpha_e(\omega)\Delta_e(\omega) \coth\left(\frac{\hbar\omega}{2k_B T}\right) \quad (37)$$

$$F_z^{(1)}(z) = -\frac{\hbar}{2\pi^2} \text{Im} \int_0^\infty d\omega \int_{-\infty}^{+\infty} dk_x \int_{-\infty}^{+\infty} dk_y \cdot \exp(-2q_0 z)\alpha_e(\omega) \cdot$$
$$\cdot [\Delta_e(\omega)(2k^2 - \omega^2/c^2) + \Delta_m(\omega)\omega^2/c^2] \cdot \coth\left(\frac{\hbar\omega}{2k_B T}\right)$$
(38)

One can see that Eq.(37) does not contain the reflection factor $\Delta_m(\omega)$, whereas coefficient $(k^2 - \omega^2/c^2)$ differs from $(2k^2 - \omega^2/c^2)$ in Eq.(38). This contradicts to the well recognized theory of the Casimir –Polder force [8,9-15], since Eq.(37) does not depend on reflection coefficient of the S-polarized electromagnetic modes.

Third, let us compare the heating rates. We claim that Eqs.(34), (36) in Ref.[20] are in error because their integrands contain incorrect frequency factor $\omega$ instead of the Doppler –shifted one, $\omega^+ = \omega + k_x V$. This error results in important physical consequence. As we have demonstrated in [16], at equilibrium $T_1 = T_2 = T$, in the lowest order velocity expansion with neglect of retardation, the total heat increment for both of the interacting plates becomes negative, that conflicts with the second law of thermodynamics. In addition to this, in Ref. [20] the factors, analogous to ours $R_{e,m}(\omega,\mathbf{k})$ in Eqs.(26),(27) are also in error. This error has been reproduced in numerous works of Volokitin et. al. since their old papers [32]. As a matter of fact, the same factors (26), (27), different from those in Refs. [20,32], come into the integrand expression of $\dot{Q}^{(1)}$ at $V=0$ [2, 35, 36], the expressions of $F_z^{(1)}$ at $V=0$ [10,13-15] and spectral density of equilibrium fluctuating electromagnetic field near a plane surface [33,34].

Moreover, as far as concerned to our relativistic expressions for $F_x^{(1)}(z), F_z^{(1)}(z), \dot{Q}^{(1)}(z)$, and their equivalents, Eqs. (22), (28) and (36) in Ref. [20], the last ones manifest a very curious mixing between S- and P- wave contributions to the forces and heat transfer due to relativistic effects, that conflicts with Eqs.(1)-(3). In total, we claim that, despite in their works [5,20,32] the authors employ a currently used basics of fluctuation electrodynamics, they have done a fatal mistake from the very beginning [32], when they calculated frictional forces in a system of thick parallel plates under thermal equilibrium. This has led to the principally incorrect expression for the force component related with propagating electromagnetic modes (the first integral term in Eq.(35)). This erroneous result has been reproduced in the following papers up to Ref. [20], where it has outcome from in the limit of small velocities. Some other errors have appeared when the authors passed from configuration 2 to configuration 1.

Finally, we briefly touch upon the recent very intriguing results obtained by Philbin et.al. [21]. These authors came to the conclusion that there is no lateral force on the plates in relative motion, as well as on a particle moving at a constant speed parallel to a plate. Similarly to [20], the problem of FEI in configuration 2 has been considered in relativistic statement, calculating the forces applied to the plates from the Maxwell stress tensor. The state and role of vacuum background have not been defined. The main results obtained in [21] for $F_x^{(2)}(z), F_z^{(2)}(z), \dot{Q}^{(2)}(z)$ have great resemblance to the results [20], however, their principal difference is that the Doppler –shifted frequency in hyperbolic cotangent of the thermal factors comes under modulus sign. This results in zero lateral force at $T_1 = T_2 = 0$ in any order of the velocity expansion and contradicts to the well substantiated results [2,5,6,8,16,17,20,25,32]. Particularly, a finite dissipative (frictional) force at $T_1 = T_2 = 0$ exactly follows from Eqs. (1) and (30). A detailed discussion of the results [21] is out a scope of this paper, and we only wish to note that, contrary to vacuum electromagnetic modes, surface electromagnetic modes of the plates at $T_1 = T_2 = 0$ do not have a relativistic invariance, so there is no rigid physical reason for the lateral force to be zero.

## 6. Conclusion

With account of our exact solution to the relativistic problem of fluctuation electromagnetic interaction in configuration 1 (a small particle moving near a wall), and using a correspondence rule between configurations 1 and 2 (two thick featureless parallel plates in relative motion), we have calculated conservative –dissipative forces and rate of heat exchange in configuration 2 in the retarded relativistic approximation of fluctuation electrodynamics at nonrelativistic relative velocity of one of the plates. Simple transition rules between both configurations are substantiated. It is shown that fluctuation electromagnetic forces and heating rates in configurations 1 and 2 can be strictly derived from one another in the limit of rarified medium for one of the plates. These results may be regarded as important referring points in future elaboration of relativistic problem in configuration 2. Nevertheless, correct solution of this problem makes it necessary to clear up an important specifics of the surrounding vacuum background. To date, therefore, a solution of the general relativistic problem in configuration 2 ($\beta \to 1, T_1 \neq T_2 \neq T_3$) still remains an open question.

Captions to figures:

Fig.1(a) Configuration 1. Geometry of motion of a particle and a Cartesian reference frame associated with the surface of the medium (system $K$). The Cartesian axes ($x', y', z'$) of the particle rest frame $K'$ are not shown.

Fig.1(b) Configuration 2, corresponding to large thick plates 1 and 2 at temperatures $T_1$ and $T_2$ in the rest frame of each one, respectively. $K$ and $K'$ are the corresponding Cartesian reference frames. The surrounding vacuum background has temperature $T_3$.

FIGURE 1(a)

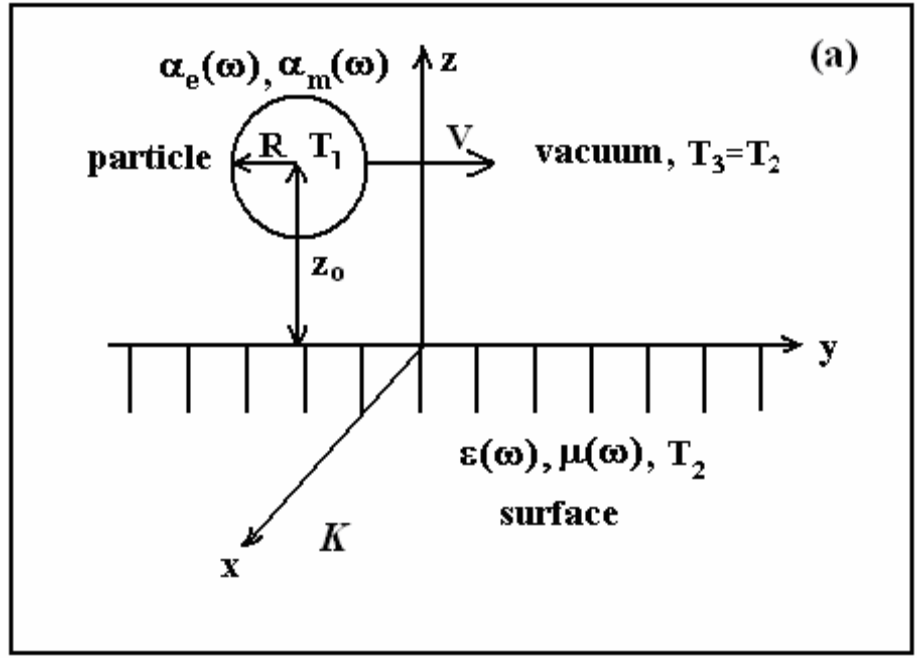

FIGURE 1(b)

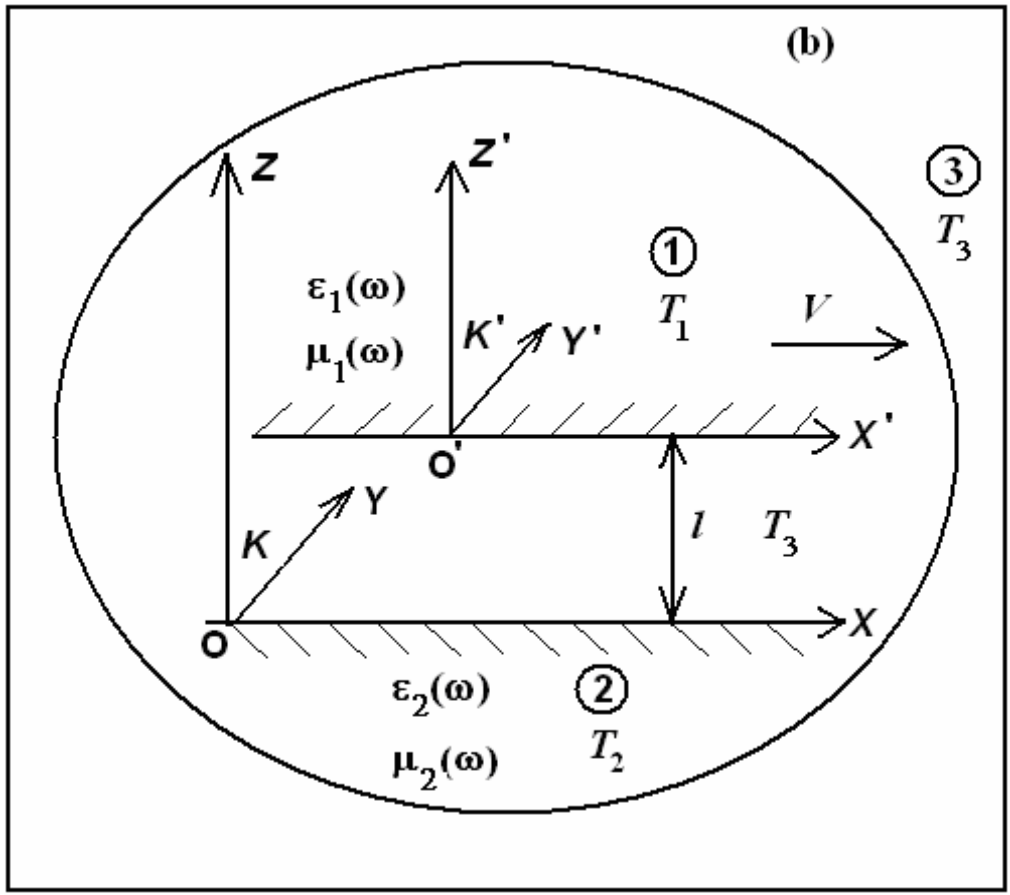